\documentclass[prd,preprint, nofootinbib,amsmath,amssymb, aps, floatfix]{revtex4-1}
\pdfoutput=1
\usepackage{dcolumn}
\usepackage{bm}

\usepackage{amssymb,amsmath,graphicx}
\usepackage[linktoc=page]{hyperref}
\usepackage{tikz}
\usepackage{fancyhdr}
\usetikzlibrary{decorations.pathmorphing}
\newcommand{\T}{\mathfrak{t}}

\begin{document}
	\begin{flushright}
		IPM/P-2018/066	
	\end{flushright}
\title{ Complexity Growth Following Multiple Shocks }
\author{M. Reza Tanhayi ${}^{1,2}$}
\email{mtanhayi@ipm.ir}
\author{ R. Vazirian$^{1}$}	
\email{s.vazirian@srbiau.ac.ir}
\author{ S. Khoeini-Moghaddam$^{3}$}
\affiliation {\small{${}^1$Department of Physics, Faculty of Basic Science, Islamic Azad
		University Central Tehran Branch (IAUCTB), P.O. Box 14676-86831,
		Tehran, Iran}}
\affiliation{\small{${}^2$School of Physics, Institute for Research in Fundamental Sciences 	(IPM) P.O. Box 19395-5531, Tehran, Iran}}
\affiliation{${}^3$
	Department of Astronomy and High Energy Physics, Faculty of Physics, Kharazmi University, Mofateh Ave., Tehran, Iran}
\vspace{1cm}	
\begin{abstract}
		
In this paper by making use of the ``Complexity=Action" proposal, we study the complexity growth after shock waves in holographic field theories. We consider both double black hole-Vaidya and AdS-Vaidya with multiple shocks geometries. We find that the Lloyd's bound is respected during the thermalization process in each of these geometries and at the late time, the complexity growth saturates to the value which is proportional to the energy of the final
state. We conclude that the saturation value of complexity growth rate is independent of the initial temperature and in the case of thermal initial state, the rate of complexity is always less than the value for the vacuum initial state such that considering multiple shocks it gets smaller. Our results indicate that by increasing the temperature of the initial state, the corresponding rate of complexity growth starts far from final saturation rate value.

\end{abstract}
\maketitle
\newpage
\tableofcontents \noindent \hrulefill

\section{Introduction}
It is claimed that the gauge/gravity duality may shed light on better understanding of the nature of spacetime by  providing a relation between  entanglement and quantum gravity (see for example Ref.s \cite{VanRaamsdonk:2009ar,VanRaamsdonk:2010pw}). In fact, with the help of this duality some certain quantum mechanical quantities can be understood classically by means of the geometry. For instance the entanglement entropy -that provides insight into quantum mechanical interpretations of the gravitational entropy- can be described holographically by minimization of an area of a codimension-two hypersurface in the bulk geometry \cite{Nishioka:2009un}. One interesting quantum mechanical object is in fact the complexity. 
In principle the complexity of quantum state which originates from the field of
quantum computations is defined by the number of elementary unitary operations which is required to build up a desired state from a given reference state \cite{Aaronson:2016vto}. It was also argued that  quantum complexity helps us to
capture some certain features of the late time behavior of eternal black hole geometries \cite{Susskind:2014moa}. \\In order to describe the complexity from the holographic point of view, there are two conjectures made by Susskind et al. The first one pointed out that the black hole interior volume at the gravity side is dual to the complexity of the boundary system at the CFT side \cite{Susskind:2014rva}. This proposal is known as the "Complexity=Volume" or CV duality. The other one was given in Ref.s \cite{Brown:2015bva}, in
which the computational complexity of a state at time $\T$ is connected to the classical bulk on-shell action in the Wheeler-DeWitt (WDW) patch. This proposal is known as the "Complexity=Action" or CA duality and is given by
\begin{equation}
{\cal C}=\frac{{\cal I}}{\pi\hbar},
\end{equation}
where ${\cal C}$ stands for the complexity of boundary state  and ${\cal I}$ is the on-shell gravitational action of region inside the WDW patch.\footnote{Subregion complexity was also defined in \cite{Alishahiha:2015rta,Ben-Ami:2016qex, Carmi:2016wjl}.}	 
Inspired by the  Lloyd's bound \cite{Lloyd},  in Ref.s  \cite{Brown:2015bva} it is argued that the rate of the complexity growth of a given state  is bounded by the average energy of the state. Lloyd's bound states that the energy in the system puts an upper limit to the rate of computation. The conjectured Lloyd's bound is given by
\begin{equation}
\frac{d}{d\T}{\cal C}(\T)\le \frac{2E}{\pi},
\end{equation}	
where $E$ is the average energy of the state at a given boundary time $\T$ (note we take $\hbar=1$). To test this conjectured bound, several works have been done and it was shown that in some certain cases in computation of the holographic complexity, the bound is violated. For example, in the case of Schwarzschild black hole which is dual to a thermofield double state, the holographic complexity violates the bound \cite{5}. It is pointed out that the two-sided eternal AdS black hole is dual to two copies of a CFT in the thermofield double state ($ |TFD\rangle$) \cite{Maldacena:2001kr}. This kind of duality provides a setup to study some aspects of black hole physics and also quantum information theory, for example  emergent spacetime \cite{VanRaamsdonk:2010pw} and ER=EPR \cite{Susskind:2014yaa}.
The special feature of such states is that by tracing out in one side results in the thermal density matrix at inverse temperature for the other side \cite{Shenker:2013pqa}. Single perturbation of thermofield double state by an operator $W(t_1)$ of the form \begin{equation}
W(t_1)\,\,|TFD\rangle,
\end{equation}
raises the energy by an amount of order the temperature of the black hole. Van Raamsdonk \cite{VanRaamsdonk:2013sza}, Shenker and Stanford \cite{Shenker:2013pqa} have argued that such perturbations on one side at enough large negative time can potentially create high energy shock waves which propagates to the other side. Namely, $n$ shock waves are constructed by perturbing $|TFD\rangle$ with $n$ thermal scale operators as follows:\begin{equation}
W_n(t_n)\cdots W(t_1)\,\,|TFD\rangle.
\end{equation}
In the shock wave geometries dual to the above perturbed thermofield double state, the complexity has been studied in Ref. \cite{Stanford:2014jda}. In this paper we use CA conjecture to examine the complexity in the presence of many shocks in both the one-sided black hole which is dual to a single CFT and AdS-Vaidya geometry.
\begin{figure}[h!]
	\centering
	\includegraphics[width=6cm]{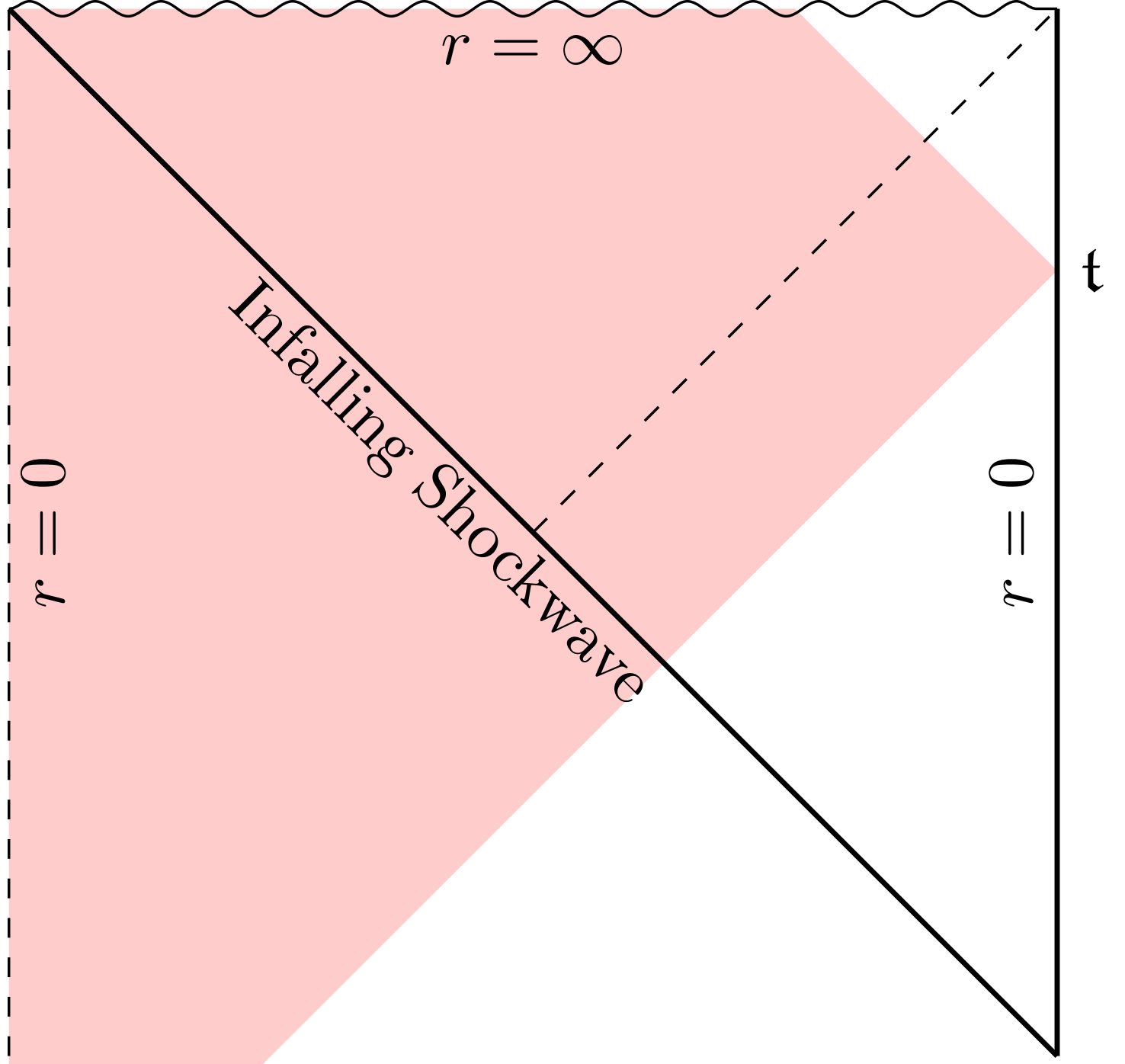}\,\,\,\,\,\,\,\,\,\,\,\,\includegraphics[width=7cm]{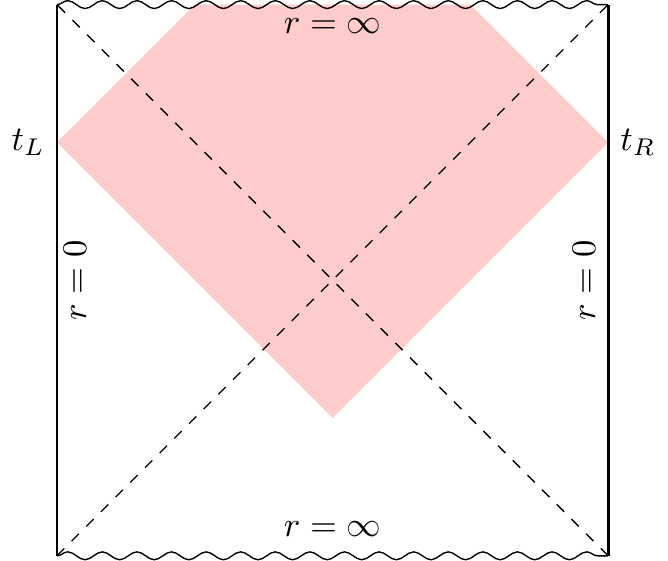}
	\caption{The Penrose diagrams for one-sided black hole (\emph{left}) and for two-sided eternal black holes (\emph{right}). The shaded region denotes
		the WDW patch corresponding to boundary times. According to the CA proposal, complexity of the CFT state corresponds to the action which should be computed inside the WDW patch \cite{Brown:2015bva}.   } 
	\label{dbh-fig}
\end{figure}


In the time dependent Vaidya spacetime, using the CA proposal, the complexity growth has been studied in \cite{Moosa:2017yvt, Alishahiha:2018tep}. More recently in \cite{Chapman:2018dem,Chapman:2018lsv}, Myers et al. investigated holographic complexity for eternal black hole backgrounds perturbed by shock waves. It was also shown that injection the shock in one side at local boundary time $\T$, raises the mass of the black hole  \cite{Chapman:2018lsv}. Time dependent Vaidya spacetime describes a collapsing thin shell of null matter with arbitrary energy to form a black hole. This geometry leads to quantum quenches of a system at the CFT side where this injection excites the system out of equilibrium and eventually the system evolves towards an equilibrium state \cite{Alishahiha:2014cwa,Tanhayi:2015cax}. \\
 The main goal of this paper is to further explore the Lloyd's bound when the gravity side undergoes the two and many shock waves. In fact in this paper we use CA proposal and generalize the work of Ref. \cite{Moosa:2017yvt} for multiple shocks and also for the black hole initial state.  \\
The rest of this paper is organized as follows. In section 2, we will consider the holographic complexity growth in a double black hole-Vaidya background. In this geometry there is a shock wave in the black hole background which follows a global quench of thermal state and despite the AdS-Vaidya background, the initial state is already thermal (see Ref. \cite{Ageev:2017wet} for more details). In section 3, we will investigate the Lloyd's bound in time dependent geometry after many shocks. Finally, a brief summary will be presented in section 4. Some details of our holographic calculations can be found in Appendix.

\section{Holographic complexity growth in a double black hole-Vaidya background }

In this section we study the holographic complexity growth after a global quench of a thermal state. Basically, in this setup the thermalization starts from non-zero temperature and after this sharp perturbation, the system thermalizes and reaches equilibrium. The holographic dual of this process is provided by an injection of a thin shell of null matter (shock wave) in the black hole background and the corresponding geometry can be well described by a double black hole-Vaidya metric. In other words, in this geometry one deals with a black hole evolving from the initial state with the horizon $r_{h_2}$ to the final state with the horizon $r_{h_1}$. Such an injection causes the temperature of the underlying system increases from its initial value $T_i$ to $T_f$. 
 In this section we study the evolution of holographic complexity for double black hole-Vaidya background; this geometry is given by

\begin{equation}\label{metric}
ds^{2} = \frac{R^2}{r^{2}} \, \Big( -f(v,r) \, dv^{2} - 2 dv dr + dx^{2}_{i}\Big),\,\,\,\,i=1,\cdots,d,
\end{equation}
where
\begin{align}
f(v,r) =
\begin{cases}
1-(\frac{r}{r_{h_2}})^{d+1}\equiv f_{2}(r) &\text{for $v <0$} \, ,\\[0ex]
1 - (\frac{r}{r_{h_1}})^{d+1}\equiv f_{1}(r) &\text{for $v>0$} \, , \label{eq-fv}
\end{cases}
\end{align}
and $R$ is a typical length scale that we set it to one. The Penrose diagram of this spacetime is shown in Fig.~(\ref{fig-2}).

In order to compute the complexity for the boundary state at time $\T$ according to the CA conjecture, we need to find the on-shell action in the corresponding WDW patch; following \cite{Moosa:2017yvt, Alishahiha:2018tep} the corresponding patch can be separated into two parts $v>0$ and $v<0$. Thus, in what follows we want to find ${\cal I}_{v>0}$ and ${\cal I}_{v<0}$ that are the gravitational action for shaded regions $v>0$ and $v<0$, respectively and the total on-shell action becomes
\begin{equation}
{\cal I} = \, {\cal I}_{(v>0)} + {\cal I}_{(v < 0)} \, , \label{eq-ac-total}
\end{equation}
since the on-shell action of a null dust vanishes.

\subsection{Calculation of action for $v>0$ region}

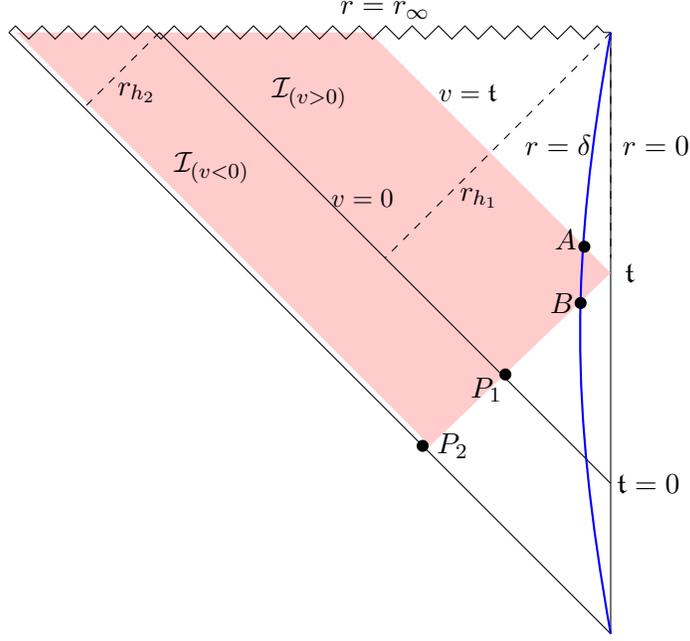
\begin{figure*}[h]
	\begin{tikzpicture}
	\draw [white](0,4.05) --(-8,4.05);
	\draw [black](0,-4.0) --(0,4);
	\fill [red!20!white](0,.8) --(-3.2,4) --(-7.9,4) --(-2.4,-1.5)--(0,.8) ;
	\node at (-4,3.2) { ${\cal I}_{(v>0)}$};
	\draw [black](0,-2.0) --(-6.0,4);
	\draw [decorate,decoration=zigzag](-8,4) --(0,4);
	\draw [dashed](0,4) --(-3,1);
	\draw [black](0,-4) --(-8,4);
	\node at (.5,-2.0) {$\T=0$};
	\node at (.25,0.8) {$\T$};
	\node at (-1.75,1.8) { {\small $r_{h_1}$}};
	\draw [dashed](0,4) --(0,1);
	\draw [dashed](-6,4) --(-7,3);
	\node at (-6.3,3.2) { $r_{h_2}$};
		\node at (-3,4.3) {$r=r_\infty$};
			\node at (.6,2.5) {$r=0$};
			\node at (-.7,2.5) {$r=\delta$};
	\node at (-1.65,-0.75) { ${P_1}$};
	\fill [black](-1.4,-.55) circle (.08cm);
	\node at (-5.3,2.2) { ${\cal I}_{(v<0)}$};
	\node at (-2.1,-1.5) { ${P_2}$};
	\fill [black](-2.5,-1.5) circle (.08cm);
	\node at (-3.3,1.8) {{\footnotesize $v=0$}};
	\node at (-1.9,3.2) {{\footnotesize $v=\T$}};
	\draw [blue,thick](0,4) to [out=260,in=100] (0,-4);
	\node at (-0.6,1.25) {{\small $A$}};
	\fill [black](-0.35,1.150) circle (.08cm);
	\node at (-0.65,0.40) {{\small $B$}};
	\fill [black](-0.4,0.40) circle (.08cm);
	
	\end{tikzpicture}
	\caption{The Penrose diagram of double black hole-Vaidya geometry. The time dependent geometry is resulted from collapse of a null shell where the injection takes place at the boundary time $\T=0$. The horizons are shown by $r_{h_1}$ and $r_{h_2}$. The WDW patch corresponding to the boundary time $\T$ for $v\gtrless0$ is shown by the shaded region which includes both inside and outside of the collapsing null shell. The cut-off surface (i.e. $r=\delta$) is denoted by blue curve. The points $A$ and $B$ are the intersections of the WDW patch with  $r=\delta$. The intersections of the past null boundary of the WDW patch and the collapsing null shell are denoted by ${P_1}$ and ${P_2}$.}
	\label{fig-2}
\end{figure*}

In this region the $r=\delta$ can be defined as a regulator of the UV divergence and hence as it is clear from Fig.~(\ref{fig-2}), there are indeed five boundaries: A spacelike boundary located at the future singularity $(r=\infty)$, a timelike boundary at $r=\delta$ and three null boundaries located at $v=0, \,\,v=\T$ and the one joining points $B$ and $P_1$. The intersections of the past null boundary of the WDW patch and the collapsing null shell are denoted by  black dots and are labeled by $P_1$ and $P_2$.
The null boundary between points $B$ and  $P_1$ which is denoted by $r_1(v,\T)$ satisfies the following integral equation \begin{equation}
\frac{1}{2}(\T - v) = \, \int_{0}^{r_1(v;\T)}  dr \, \frac{1}{f_{1}(r)}.
\end{equation}
Inserting the explicit form of the $f_{1}$ results in
\begin{equation}\label{eq7}
\T-v=2r_{1}(\T)\,\,\,{}_2F_1\Big(1,\frac{1}{d+1},1+\frac{1}{d+1},(\frac{r_{1}}{r_{h_1}})^{d+1}\Big).
\end{equation}

The coordinates of the points $A$, $B$, and $P_1$ are
\begin{align}
A: \,\,\,\, v_{A} =& \, \T \, , \,\,\,\,\,\,\,\,\,\,\,\,\,\,\,\,\,\,\,\,\,\,\,\,\, r_{A} = \, \delta \, ,\label{eq-cor-A} \\
B: \,\,\,\, v_{B} =& \, \T - 2\delta \, , \,\,\,\,\,\,\,\,\,\,\,\, r_{B} = \, \delta \, , \label{eq-cor-B}\\
P_1: \,\,\,\, v_{P_1} =& \, 0 \, , \,\,\,\,\,\,\,\,\,\,\,\,\,\,\,\,\,\,\,\,\,\,\,\, r_{P_1}(\T) = \, r_{1}(0,\T) \, . \label{eq-cor-z}
\end{align}
It is noted that the coordinate for $B$ is obtained by expanding Eq. \eqref{eq7} around ${r_P} \sim \delta$ that leads to
\begin{equation}
\T-v_B=2\delta \lim_{\delta\rightarrow0}{}_2F_1 \sim 2\delta.
\end{equation}
The bulk contribution to the gravitational action is given by the Einstein-Hilbert term. However, the Ricci scalar is constant in this region and the bulk contribution to ${\cal I}_{(v>0)}$ becomes proportional to the volume of the corresponding region. Thus, we have
\begin{align}
{\cal I}_{(v>0)}^{\text{bulk}} &= \frac{1}{16\pi G_N}\int\sqrt{-g} \Big( R - 2 \Lambda \Big)d^{d+2}x\nonumber \\&=\frac{1}{16\pi G_N} \bigg(\, \int d^{d}x \, \int_{0}^{\T-2\delta} dv \, \int_{r_{1}(v;\T)}^{\infty}  \, \frac{dr}{r^{d+2}} \, +  \, \int d^{d}x \, \int_{\T-2\delta}^{\T} dv \, \int_{\delta}^{\infty}  \, \frac{dr}{r^{d+2}} \, \bigg) \nonumber\\
&= -\frac{V_d}{8\pi G_N}\,\Big(\frac{2}{\delta^{d}}  + \int_{0}^{\T-2\delta}  \, \frac{dv}{r^{d+1}_{1}(v;\T)}\Big) \, ,
\end{align}
where we have defined
\begin{equation}
V_d \equiv \int d^{d}x \, , \label{eq-size}
\end{equation}
as the volume of $d$-dimensional boundary system parametrized by $x_i,\,\,\,i=1,\cdots,d$. Also $G_N$ is the Newton's constant and $R$ and $\Lambda$ are respectively the Ricci scalar and cosmological constant given by
\begin{equation}
R=-(d+1)(d+2),\,\,\,\,\,\,\Lambda=\frac{-1}{2}d(d+1).
\end{equation}

Now we should compute the boundary (surface) contributions to ${\cal I}_{(v>0)}$. 
The surface gravity of the null boundaries would be vanished by using the affine parametrization for the null directions, so these null boundaries have no contribution in computing the on-shell action. On the other hand, for the timelike boundary there is no time dependency in the corresponding surface term. Since our ultimate goal is to compute the time derivative of the complexity,  we ignore the calculation of the boundary term for the timelike surface. \\
Therefore, we have to consider the spacelike surface at $r=r_\infty$. The contribution of this boundary is given by Gibbons-Hawking term. To compute this term, we first consider the spacelike surface at $r = r_{\infty} \gg r_{h_1}$ and then take the limit $r_{\infty} \to \infty$. The future-directed normal vector to the $r = r_{\infty}$ surface is
\begin{equation}
n^{i} = \, -\frac{r_{h}^{\frac{d+1}{2}}}{r_{\infty}^{\frac{d-1}{2}}} \, \big(\partial_{v}\big)^{i} - \frac{r_{\infty}^{\frac{d+1}{2}+1}}{r_{h}^{\frac{d+1}{2}}} \, \big(\partial_{r}\big)^{i} \, .
\end{equation}
Thus, the corresponding Gibbons-Hawking term is given by
\begin{equation}
{\cal I}_{(v>0)}^{(r=\infty)} = \, \lim_{r_{\infty}\to \infty} \, - \frac{1}{8\pi G_N} \,  \int d^{d}x \, \int_{0}^{\T} dv \, \sqrt{|\widetilde{\gamma}|} \, \, \widetilde{\gamma}^{ij}\nabla_{i}n_{j} \, ,\label{eq-ghy-2}
\end{equation}
where $\widetilde{\gamma}^{ij} \equiv g^{ij} + n^{i}n^{j}$ is the inverse induced metric on the $r=r_{\infty}$ surface. The negative sign in Eq.~\eqref{eq-ghy-2} is due to the fact that the shaded region is to the past of the $r=r_{\infty}$ boundary.  Solving the integrals in Eq.~\eqref{eq-ghy-2} yields
\begin{equation}
{\cal I}_{(v>0)}^{(r=\infty)} = \, \frac{V_d}{16\pi G_N} \, \frac{d+1}{r_{h_1}^{d+1}} \, \T \, .
\end{equation}

Finally, we consider the contributions of the joint points where a null boundary intersects with another boundary. These points are located at $A$, $B$ and $P_1$ and their contributions to the on-shell action are given by \cite{Lehner:2016vdi}
\begin{align}
{\cal I}_{(v>0)}^{\text{joint}} =& \, -\frac{1}{8\pi G_N} \int_{A} d^{d}x \, \sqrt{\gamma_{\text{ind}}} \, \log \left|k_{\text{in}}\cdot s\right|  - \, \frac{1}{8\pi G_N} \int_{B} d^{d}x \, \sqrt{\gamma_{\text{ind}}} \, \log \left|k_{\text{out}}\cdot s\right| \nonumber\\&+ \, \frac{1}{8\pi G_N} \int_{P_1} d^{d}x \, \sqrt{\gamma_{\text{ind}}} \, \log \left|\frac{ k_{\text{in}}\cdot k_{\text{out}}}{2}\right| \, ,
\end{align}
where $\sqrt{\gamma_{\text{ind}}} = r^{-d}$ is the determinant of the induced metric of the codimension-two corners,
the $s^i$ stands for the normal vector of the cut-off boundary $(r=\delta)$ which is given by
\begin{equation}
s^i=\delta (\partial_{v})^i-\delta(\partial_{r})^i.
\end{equation}
and $k_{\text{in}}$ and $k_{\text{out}}$ are the null generators which are
\begin{align}
k_{\text{in}}^{i} =& \, -\alpha r^{2} \, \big( \partial_{r}\big)^{i} \, ,\label{eq-kin}\\
k_{\text{out}}^{i} =& \, \beta\Big(\frac{2 r^{2}}{f_{1}(r)} \, \big( \partial_{v}\big)^{i} - r^{2} \, \big( \partial_{r}\big)^{i}\Big) \, .
\end{align}
It is straightforward to show that the joint terms resulted from $A$ and $B$ are in fact time-independent, thus it is only needed to compute the joint term coming from the $P_1$.
Therefore, one obtains
\begin{equation}
{\cal I}_{(v>0)}^{\text{joint}} =  \frac{V_d}{8\pi G_N} \, \frac{1}{r_{P_1}^{d}(\T)} \Bigg(2 \log r_{P_1}(\T) - \log f_{1}\big(r_{P_1}(\T)\big)+\log (\alpha\beta\big) \Bigg) \, .
\end{equation}

Thus, the corresponding gravitational action for $v>0$ region of Fig.~(\ref{fig-2}) is given by the sum of the contributions from the bulk, surfaces and joint terms which is given by
\begin{align}\label{26}
{\cal I}_{(v>0)}= &-\frac{V_d}{8\pi G_N}\,\int_{0}^{\T-2\delta}  \, \frac{dv}{r^{d+1}_{1}(v;\T)}\nonumber\\ &+\frac{V_d}{16\pi G_N} \, \frac{d+1}{r_{h_1}^{d+1}} \, \T +\frac{V_d}{8\pi G_N} \, \frac{1}{r_{P_1}^{d}(\T)} \Bigg(2 \log r_{P_1}(\T) - \log f_{1}\big(r_{P_1}(\T)\big)+\log (\alpha\beta\big) \Bigg),
\end{align}
noting that we have dropped the time-independent terms.

\subsection{Calculation of action for $v<0$ region}

In this region there are three boundaries. One of them is the spacelike boundary at $r=r_\infty$, the other is the in-falling null shell at $v=0$ and the third one is the null boundary that connects points $P_1$ and $P_2$ denoted by $r_2(v,\T)$ which is satisfying the following integral equation
\begin{equation}\label{27}
v=2\int_{r_2(v,\T)}^{r_{P_1}(\T)}\frac{dr}{f_{2}(r)}.
\end{equation}
Similar to the $v>0$ region, the bulk contribution to ${\cal I}_{(v<0)}$ is given by the Einstein-Hilbert term which is indeed proportional to the volume of the shaded region for $v<0$ in Fig.~(\ref{fig-2}). It is given by
\begin{equation}
{\cal I}_{(v<0)}^{\text{bulk}} = \, \frac{1}{16\pi G_N} \, \int d^{d}x \, \int_{-\infty}^{0} dv \, \int_{r_2(v,\T)}^{\infty} ( R - 2 \Lambda)\frac{dr}{r^{d+2}},
\end{equation}
which leads to
\begin{align}
{\cal I}_{(v<0)}^{\text{bulk}} =-\frac{V_d}{8\pi G_N}\int_{-\infty}^{0}\frac{dv}{r_2^{d+1}(v,\T)},
\end{align}
where we have used  $R-2\Lambda = -2(d+1)$.

We now consider the boundary contributions to ${\cal I}_{(v<0)}$. 
As long as the surface gravity of the null generators vanishes, they do not contribute to our calculations. On the other hand, the contribution of the spacelike surface located at $r=r_\infty$ would be time-independent so that we do not take its effect into account in our calculation. Therefore, we only have to focus on the null counterterms. The contribution from the counterterms at $P_1$ is given by\footnote{The corner $P_2$ has no contribution because the volume density of the codimension-two surface falls off as $r^{-d}$.   }
\begin{equation}
{\cal I}_{(v<0)}^{\text{joint}} =-\frac{1}{8\pi G_N}\int_{P_1}\sqrt{\gamma_{in}}\,d^dx\log|\frac{ k_{\text{in}}\cdot k_{\text{out}}}{2}|,
\end{equation}
where in this case the null generator $k_{\text{in}}^{i}$ is given by Eq. \eqref{eq-kin} and also one has
\begin{align}
k_{\text{out}}^{i} =& \, \beta\Big(\frac{2 r^{2}}{f_{2}(r)} \, \big( \partial_{v}\big)^{i} - r^{2} \, \big( \partial_{r}\big)^{i}\Big).
\end{align}
Therefore, one obtains
\begin{equation}\label{joint}
{\cal I}_{(v<0)}^{\text{joint}} = - \frac{V_d}{8\pi G_N} \, \frac{1}{r_{P_1}^{d}(\T)} \Bigg(2 \log r_{P_1}(\T) - \log f_{2}\big(r_{P_1}(\T)\big)+\log (\alpha\beta\big) \Bigg) \, .
\end{equation}
Putting the results together, the time dependent parts of the gravitational action for the present region becomes
\begin{equation}
{\cal I}_{(v<0)} = -\frac{V_d}{8\pi G_N}\int_{-\infty}^{0}\frac{dv}{r_2^{d+1}(v,\T)}- \frac{V_d}{8\pi G_N} \, \frac{1}{r_{P_1}^{d}(\T)} \Bigg(2 \log r_{P_1}(\T) - \log f_{2}\big(r_{P_1}(\T)\big)+\log (\alpha\beta\big) \Bigg).
\end{equation}

\subsection{On-shell action and complexity growth}

The total gravitational action of the WDW patch for the double black hole-Vaidya geometry is given by
\begin{align}
{\cal I}={\cal I}_{(v>0)}+{\cal I}_{(v<0)}=& -\frac{V_d}{8\pi G_N}\,\Big( \int_{0}^{\T-2\delta}  \, \frac{dv}{r^{d+1}_{1}(v,\T)}+\int_{-\infty}^{0}\frac{dv}{r_2^{d+1}(v,\T)}\Big)+\frac{V_d}{16\pi G_N} \, \frac{d+1}{r_{h_1}^{d+1}} \, \T\nonumber\\
&-\frac{V_d}{8\pi G_N}\frac{1}{r_{P_1}^d(\T)}\log\frac{f_{1}(r_{P_1}(\T))}{f_{2}(r_{P_1}(\T))}+(\text{$\T-$independent terms}).
\end{align}
The important fact is that since we have used the same free parameters in writing the null vectors for both $v>0$ and $v<0$ regions, the total action is independent of $\alpha$ and $\beta$.\footnote{In general, different free parameters can be used, but by adding proper counterterms, it can be shown that the final result will not change. }
\\The time derivative of the above expression gives us
\begin{align}\label{key}
\frac{d}{d\T}{\cal C}=\frac{1}{\pi}\frac{d}{d\T}{\cal I}&=\frac{{2E}}{\pi }\left[ {1 - \frac{{d + 1}}{{2d}}\frac{{{r_{{h_1}}}^{d + 1}}}{{{r_{{h_2}}}^{d + 1}}}\frac{{{f_1}\left( {{r_{{P_1}}}} \right)}}{{{f_2}\left( {{r_{{P_1}}}} \right)}} + \frac{{{f_1}\left( {{r_{{P_1}}}} \right)}}{2}\frac{{{r_{{h_1}}}^{d + 1}}}{{{r_{{P_1}}}^{d + 1}}}\log \frac{{{f_1}\left( {{r_{{P_1}}}} \right)}}{{{f_2}\left( {{r_{{P_1}}}} \right)}}} \right]
\end{align}
where $E$ is the energy of the equilibrium state in the boundary which is given  in terms of the radius of the last horizon as follows
\begin{equation}
 E = \, \frac{V_d}{16\pi G_N} \, \frac{d}{r_{h_1}^{d+1}},
\end{equation}
and also the corresponding temperature is $T = \, \frac{d+1}{4\pi\, r_{h_1}}$. In fact, the expression \eqref{key} gives us the bound on the rate of complexity. It is worth mentioning that since $r_{P_1}$ never crosses the event horizon $ r_{h_1}$ (i.e. ${r_{{h_1}}}>{r_{{P_1}}})$, and also $r_{h_2}>r_{h_1}$, it can be concluded that the second and third terms on the right side
of Eq. \eqref{key} are always negative. Consequently, one can say
\begin{equation}
\frac{d}{d\T}{\cal C}\le \frac{2E}{\pi},
\end{equation}
which is in agreement with the Lloyd's bound. On the other hand, at the late time $\T\gg r_{h_1}$ which means $r_{P_1}\rightarrow r_{h_1}$, the rate of complexity growth eventually saturates and its saturation value becomes $\frac{2E}{\pi}$. \\ 
It is worth mentioning that in Eq. \eqref{key} by taking the limit of $r_{h_2}$ tends to the null infinity one gets the AdS-Vaidya geometry and in this limit \eqref{key} turns to 
\begin{equation}\label{moosa}
\frac{d}{d\T}{\cal C}\sim\frac{{2E}}{\pi }\left[ {1 + \frac{1}{2}\left( {\frac{{r_{{h_1}}^{d + 1}}}{{r_{{P_1}}^{d + 1}}} - 1} \right)\log {f_1}({r_{{P_1}}})} \right].
\end{equation} 
With $E$ being the average energy of the state at time $\T$, the above equation  is the same as already derived in  \cite{Moosa:2017yvt}. On the other hand turning off the shock at time $\T=0$ (or equivalently $r_{h_2}\rightarrow r_{h_1}$) results in 
\begin{equation}
\frac{d}{dt}{\cal C}\sim\frac{d-1}{d}\frac{E}{\pi},
\end{equation}
one observes that the above growth rate is lower than the late time  saturation value.\footnote{Actually the Lloyd's bound is respected in this case by considering proper counterterms \cite{Chapman:2018dem}. We thank the referee for his/her useful comment on this point.}

\section{Complexity growth after shock waves}
In this section we are interested in the rate of complexity in a strongly coupled CFT after a sudden change. At the gravity side, despite the previous section, the initial geometry is supposed to be AdS. In the case of one shock, the complexity growth has been studied in \cite{Moosa:2017yvt}.  It was shown that the rate of complexity growth saturates the bound soon after the system reaches local equilibrium and the result is given by \eqref{moosa}. In this case, one gets $\frac{d}{d\T}{\cal C}\le \frac{2E}{\pi}$
and the saturation takes place at late time when $r_{P_1}\rightarrow r_{h_1}$. In what follows, we extend this consideration for two and more shock waves.
\begin{figure}[h!]
	\centering
	\includegraphics[width=8cm]{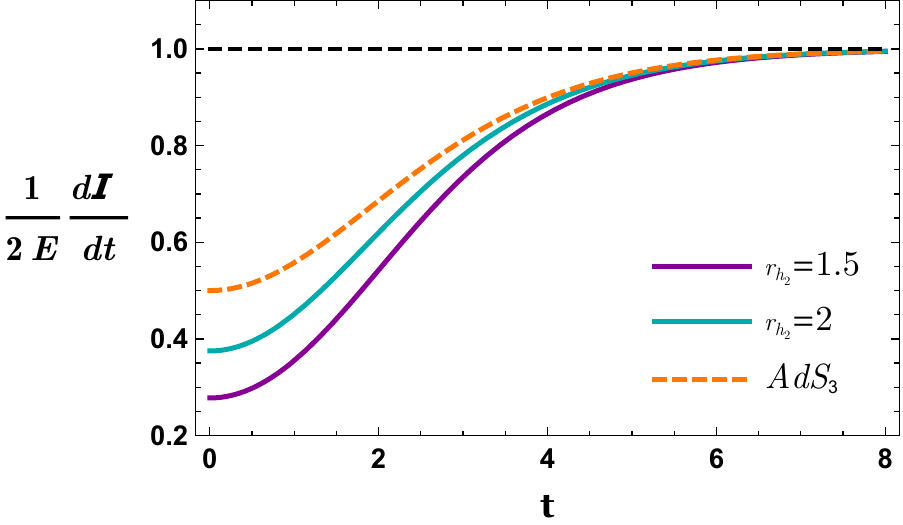}\,\,\,\,\,\,\,\,\,\,\,\,\includegraphics[width=8cm]{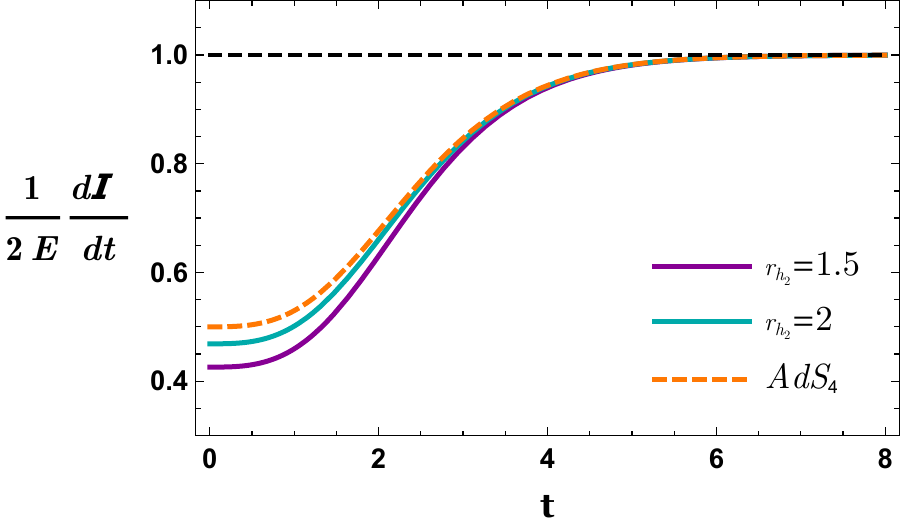}
	\caption{Complexity growth rate in AdS- and double black hole-Vaidya geometries for $d=2$ (\emph{left}) and $d=3$ (\emph{right}). In both of the plots ${r_{{h_1}}}$ is set to one. It is clear that by increasing the temperature of the initial state (i.e., for smaller $r_{h_2}$), the rate of complexity growth initiates far from final saturation value.}
	\label{dbh-fig}
\end{figure}

\subsection{Two shocks}

Let us consider black hole-Vaidya with two in-falling shells. At the field theory side, this means the system deals with energy injection twice in the conformal field theory. After first injection the system thermalizes and after the second one, the system gets more energy and hence becomes more complicated. Here, we want to study the rate of complexity due to two quenches.  \\
The metric is still given by \eqref{metric} where $f(v,r)$ is now defined as follows (we set the AdS radius to one)
\begin{align}
f(v,r) =
\begin{cases}
1- (\frac{r}{r_{h_1}})^{d+1} &\text{for $v >\T_0$} \, ,\\[0ex]
1-(\frac{r}{r_{h_2}})^{d+1} &\text{for $\T_1<v <\T_0$} \, ,\\[0ex]
1  &\text{for $v<\T_1$} \,  \label{eq-fvz2}
\end{cases}
\end{align}
In this case we have three regions: two black holes and one AdS; we are going to calculate the on-shell action in each region. The mathematical details are almost the same as previous section, so we only write the on-shell action in each region.
\begin{itemize}
	\item Region one, $v>0$:\\ The corresponding on-shell action for region one is already computed in Eq. \eqref{26} and is given by
	\begin{align}
	{\cal I}_{1}=& -\frac{V_d}{8\pi G_N}\,\int_{0}^{\T-2\delta}  \, \frac{dv}{r^{d+1}_{1}(v;\T)}\nonumber\\ &+\frac{V_d}{16\pi G_N} \, \frac{d+1}{r_{h_1}^{d+1}} \, \T +\frac{V_d}{8\pi G_N} \, \frac{1}{r_{P_1}^{d}(\T)} \Bigg(2 \log r_{P_1}(\T) - \log f_{1}\big(r_{P_1}(\T)\big)+\log (\alpha\beta\big) \Bigg).
	\end{align}
	\item Region two, $\T_1<v<0$:\\ In this region there is an additional bulk contribution which is
	\begin{align}
	{\cal I}_{(v<0)}^{\text{bulk}} =-\frac{V_d}{8\pi G_N}\int_{\T_1}^{0}\frac{dv}{r_2^{d+1}(v,\T)}.
	\end{align}
	To obtain the contribution due to the joint terms we note that there are two corners $P_1$ and $P_2$ which have almost the same contribution with different signs.\footnote{To check the sign of each terms in the action see Ref. \cite{Lehner:2016vdi}.} Thus, one can show
	\begin{align}
	{\cal I}_{2} = -\frac{V_d}{8\pi G_N}\int_{\T_1}^{0}\frac{dv}{r_2^{d+1}(v,\T)}&- \frac{V_d}{8\pi G_N} \, \frac{1}{r_{P_1}^{d}(\T)} \Bigg(2 \log r_{P_1}(\T) - \log f_{2}\big(r_{P_1}(\T)\big)+\log (\alpha\beta\big) \Bigg)\nonumber\\
	&+ \frac{V_d}{8\pi G_N} \, \frac{1}{r_{P_2}^{d}(\T)} \Bigg(2 \log r_{P_2}(\T) - \log f_{2}\big(r_{P_2}(\T)\big)+\log (\alpha\beta\big) \Bigg)
	\end{align}
	\item Region three, $v<\T_1$: \\
	It is noted that the Poincare horizon does not contribute to the on-shell action in this region, therefore, it is straightforward to check that the following expression is indeed the on-shell action in this region
	\begin{align}
	{\cal I}_{3} = -\frac{V_d}{8\pi G_N}\frac{1}{r^d_{P_2}(\T)}\bigg(\frac{2}{d}+2\log(r_{P_2}(\T))+\log(\alpha\beta)\bigg).
	\end{align}
\end{itemize}
We already have all ingredients to write the total gravitational action for this case which becomes
\begin{align}
{\cal I}=&{\cal I}_1+{\cal I}_2+{\cal I}_3=\nonumber\\
&-\frac{V_d}{8\pi G_N}\,\int_{0}^{\T-2\delta}  \, \frac{dv}{r^{d+1}_{1}(v;\T)}+\frac{V_d}{16\pi G_N} \, \frac{d+1}{r_{h_1}^{d+1}} \, \T -\frac{V_d}{8\pi G_N} \, \frac{1}{r_{P_1}^{d}(\T)} \log f_{1}\big(r_{P_1}\big)\nonumber\\
&-\frac{V_d}{8\pi G_N}\int_{\T_1}^{0}\frac{dv}{r_2^{d+1}(v,\T)}+ \frac{V_d}{8\pi G_N} \, \frac{1}{r_{P_1}^{d}(\T)}  \log f_{2}\big(r_{P_1}\big)- \frac{V_d}{8\pi G_N} \, \frac{1}{r_{P_2}^{d}(\T)}  \log f_{2}\big(r_{P_2}\big)\nonumber\\
& -\frac{V_d}{4\pi G_N}\frac{1}{r^d_{P_2}(\T)}\frac{1}{d}+\mbox{$\T$-independent terms}.
\end{align}
The time derivative of the above expression leads to 
\begin{align}
\frac{d}{d\T}{\cal C}=\frac{1}{\pi}\frac{d}{d\T}{\cal I}=\frac{2E}{\pi}\bigg[1+\frac{1}{2}f_{1}(r_{P_1}) r_{h_1}^{d+1}\bigg(\frac{1}{r_{P_1}^{d+1}} \log \frac{f_{h_1}(r_{P_1})}{f_{2}(r_{P_1})}
+\frac{1}{r_{P_2}^{d+1}}\frac{f_{2}(r_{P_2})}{f_{2}(r_{P_1})}\log f_{2}(r_{P_2})\bigg)
\bigg],
\end{align}
where the energy of the final black hole is denoted by $E$ which is the value that the rate of complexity obtained above, finally saturates to at late time i.e. $r_{P_1}\rightarrow r_{h_1}$. As long as $r_{P_1}\le r_{P_2}\le r_{h_1}\le r_{h_2}$, one can check that in this case the log terms are always negative which means the Lloyd's bound is also satisfied.\\
In the following we will generalize the discussion to $n$ collapsing null shells.

\subsection{Multiple shocks }

In this subsection we would like to study the complexity evolution after many collapsing null mass shells. Clearly, we examine the rate of change of the complexity after $n$ shocks. In this geometry, $f(v,r)$ in the metric \eqref{metric} takes the following form 
\begin{align}
f(v,r) =
\begin{cases}
1- (\frac{r}{r_{h_1}})^{d+1} &\text{for $v >\T_0$} \, ,\\[0ex]
1-(\frac{r}{r_{h_2}})^{d+1} &\text{for $\T_1<v <\T_0$} \, ,\\\hspace{.5cm}\colon\,\,\,\,\,\,\,\,\,\,\,\,\\[0ex]
1-(\frac{r}{r_{h_n}})^{d+1} &\text{for $\T_{n-1}<v <\T_{n-2}$} \, ,\\[0ex]
1  &\text{for $v<\T_{n-1}$} \,
\end{cases}
\end{align}
\begin{figure}[h!]
	\centering
	\includegraphics[width=7cm]{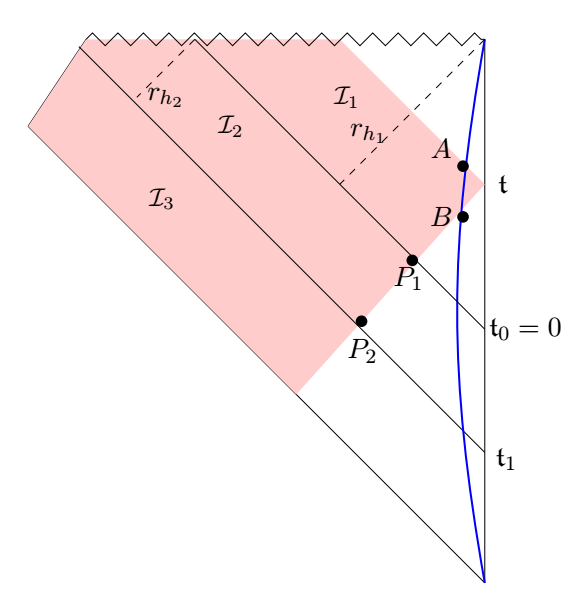}\,\includegraphics[width=8cm]{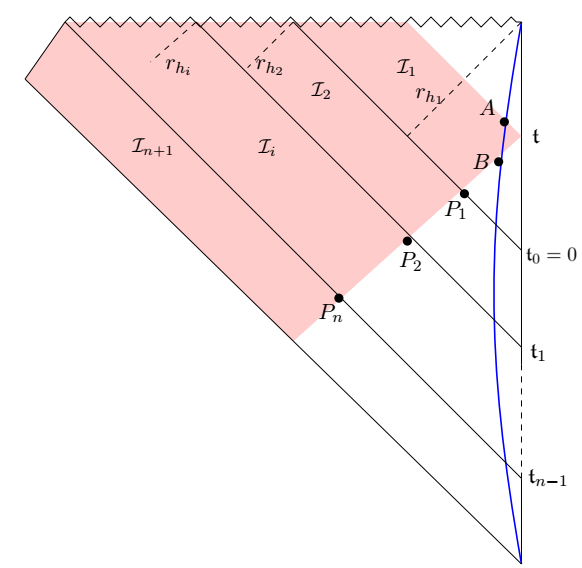}
	\caption{The Penrose diagram of AdS-Vaidya solution with two (\textit{left plot}) and $n$  (\textit{right plot}) in-falling shells. The collapsing null shells take place at boundary times $\T_0,\cdots \T_{n-1}$. The shaded region denotes the WDW patch corresponding to boundary time $\T$. The intersections of the past null boundary of the WDW patch and the collapsing shells are denoted by $P_1$, $P_2$, $\cdots$, $P_n$, noting that $r_{h_j},\,\,\,j=1,\cdots, n$ stand for the radius of the horizon in each case.  }
	\label{mutual4}
\end{figure}
In this case we are dealing with $n+1$ patches, hence the total action is given by
\begin{equation}
{\cal I} = {\cal I}_{1} + \sum\limits_{i = 2}^n {{\cal I}_{i}}  + {\cal I}_{n+1},
\end{equation}
where one can write
\begin{align}
{\cal I}_{1}=& -\frac{V_d}{8\pi G_N}\,\int_{0}^{\T-2\delta}  \, \frac{dv}{r^{d+1}_{1}(v;\T)}\nonumber\\ &+\frac{V_d}{16\pi G_N} \, \frac{d+1}{r_{h_1}^{d+1}} \, \T +\frac{V_d}{8\pi G_N} \, \frac{1}{r_{P_1}^{d}(\T)} \Bigg(2 \log r_{P_1}(\T) - \log f_{1}\big(r_{P_1}(\T)\big)+\log (\alpha\beta\big) \Bigg),
\end{align}
and the on-shell action for the region $i$ ($1<i\le n$) is given by
\begin{align}
{\cal I}_{i} = -\frac{V_d}{8\pi G_N}\int_{\T_{i-1}}^{\T_{i-2}}\frac{dv}{r_i^{d+1}(v,\T)}&- \frac{V_d}{8\pi G_N} \, \frac{1}{r_{P_{i-1}}^{d}} \Bigg(2 \log r_{P_{i-1}} - \log f_{i}\big(r_{P_{i-1}}\big)+\log (\alpha\beta\big) \Bigg)\nonumber\\
&+ \frac{V_d}{8\pi G_N} \, \frac{1}{r_{P_i}^{d}} \Bigg(2 \log r_{P_i} - \log f_{{i}}\big(r_{P_i}\big)+\log (\alpha\beta\big) \Bigg),
\end{align}
and finally, for the AdS part one has
	\begin{align}
{\cal I}_{n+1} = -\frac{V_d}{8\pi G_N}\frac{1}{r^d_{P_n}}\bigg(\frac{2}{d}+2\log(r_{P_n})+\log(\alpha\beta)\bigg).
\end{align}
To summarize, by making use of the CA conjecture, the following expression is achieved to study the bound on the complexity growth in the case of the $n$ shocks:
\begin{align} \label{fin}
\frac{d}{d\T}{\cal C}=\frac{{2E}}{\pi }\left[ {1 + \sum\limits_{j = 1}^n {\frac{{r_{{h_1}}^{d + 1}}}{{r_{{P_j}}^{d + 1}}}} \left( {\log \frac{{{f_j}({r_{{P_j}}})}}{{{f_{j + 1}}({r_{{P_j}}})}}} \right)\frac{{d{r_{{P_j}}}}}{{d\T}}} \right],
\end{align}
where in our notation ${f_{n + 1}}\left( r \right) = 1$. Similar to the case of two shocks, $E$ is the energy of the final state and in deriving Eq. \eqref{fin}, we make use of the following relation:
\begin{align}\label{k1}
\frac{dr_{P_j}}{d\T} =
\begin{cases}
\frac{{{f_1}({r_{{P_1}}})}}{2} &\text{for $j=1$} \, ,\\[0ex]
\frac{f_j(r_{P_j})}{f_j(r_{P_{j-1}})}\frac{dr_{P_{j-1}}}{d\T} &\text{for $2 \le j \le n$} \, .
\end{cases}
\end{align}
It is easy to check that in this case one also has $\frac{d}{d\T}{\cal C}\le \frac{2E}{\pi}$ and the late time saturation value becomes $\frac{2E}{\pi}$.\\
It is argued that perturbing the black hole can provide a proper setup to test the relationship between geometry and complexity \cite{Shenker:2013yza, Roberts:2014isa}. At the field theory side when the states are perturbed by a small thermal-scale operator, the perturbation can grow due to the butterfly effect and consequently the complexity of states increases. At the gravity side and in the bulk the perturbation results in an ingoing null shock wave and hence the corresponding action of the Einstein Rosen bridge increases as well.\footnote{Although in the one-sided black hole there is no second side, nevertheless one may think of the interior as a bridge leading to the same feature as the two-sided black hole \cite{Susskind:2014rva}.} Therefore, multiple shock wave states can provide more detailed evidence for
the duality between complexity and geometry; In fact one can find a well defined strategy to match geometry of tensor network  and the geometry of inside the Einstein-Rosen bridge \cite{Brown:2015lvg}. Moreover, for more than one shock wave, one expects more complexification, nevertheless the rate of complexity growth respects the Lloyd's bound as it is clear from equation \eqref{fin}.

\section{Conclusions}
Previously, Susskind and Stanford studied the holographic complexity of the thermofield double states in a wide variety of spherically symmetric shock wave geometries  \cite{Stanford:2014jda}. They conjectured that the complexity is proportional to the regularized (spatial) volume of the largest codimension-one surface crossing the Einstein-Rosen bridge. This is in fact the Complexity equals Volume proposal. In this paper, by making use of the Complexity equals Action proposal, we have studied the complexity growth rate of the conformal filed theory state after two and many shocks. We have considered both thermal and vacuum states of the CFT whose holographic dual processes can be provided by the injection of null mass shell in the black hole and AdS backgrounds, respectively.  We have shown that the conjectured Lloyd's bound is satisfied in each of the cases and notably at the late time the complexity growth saturates to its final value which is given by $\frac{2E}{\pi}$ that is independent of the initial temperature of the system.

It is worth mentioning that at the early time after the injection of null matter,  an estimation can be obtained for complexity growth which for thermal initial state from Eq. \eqref{key} becomes
\begin{align}
\frac{d}{d\T}{\cal C}\sim\frac{2E}{\pi}\,\left[ {\frac{1}{2} - \frac{1}{{2d}}\frac{{r_{{h_1}}^{d + 1}}}{{r_{{h_2}}^{d + 1}}} - \frac{1}{{{2^{d + 3}}}}\left( {\frac{1}{{r_{{h_2}}^{d + 1}}} - \frac{1}{{r_{{h_1}}^{d + 1}}}} \right)\left( {1 + \frac{{2 + d}}{d}\frac{{r_{{h_1}}^{d + 1}}}{{r_{{h_2}}^{d + 1}}}} \right){\T^{d + 1}}} \right].
\end{align}
Also in the case of the vacuum initial state one obtains
\begin{align}
\frac{d}{d\T}{\cal C}\sim\frac{2E}{\pi}\,\bigg(\frac{1}{2}+\frac{1}{2^{d+3}}\frac{\T^{d+1}}{r_{h_1}^{d+1}} \bigg),
\end{align}
which means at early time  just after the quench the complexity grows as $\T^{d+1}$.

We have also shown that the  rate of complexity evolution after a shock in the case of thermal initial sate is always less than the similar rate for the vacuum initial state and by multi shocks it gets smaller. Our results indicate that by decreasing $r_{h_2}$  which equals to higher temperature initial state, the complexity rate starts from the value which is far from its final saturation value.

In this paper we have studied quenches with the null mass shell and it would be interesting to consider massive  or charged in-falling shells.
\section*{Acknowledgments }

We would like to thank Mohsen Alishahiha for his very kind and generous
support and also for his edifying contribution to the content. We would like to acknowledge  M. Reza Mohammadi for his useful comments. We also thank A. Akhavan, A. Naseh, F. Omidi, M. Vahidinia and Mostafa Tanhayi  for some related discussions.  This work has been supported in parts by Islamic Azad University Central Tehran
Branch.

\section*{Appendix: Some useful formula}
In this appendix some details of calculation are given. Some details of the time derivative for $v>0$ region is presented below:
\begin{align}
\frac{d}{d\T}\int_{0}^{\T-2\delta}  \, \frac{dv}{r^{d+1}_{1}(v,\T)}&=\int_{0}^{\T-2\delta}\frac{d}{d\T}\frac{1}{r^{d+1}_{1}(v,\T)} dv+\frac{1}{r^{d+1}_{1}(\T-2\delta,\T)}\nonumber\\&=-\int_{0}^{\T-2\delta}\frac{d}{dv}\frac{1}{r^{d+1}_{1}(v,\T)} dv+\frac{1}{r^{d+1}_{1}(\T-2\delta,\T)}\nonumber\\
&=-\frac{1}{r^{d+1}_{1}(\T-2\delta,\T)}+\frac{1}{r^{d+1}_{1}(0,\T)}+\frac{1}{r^{d+1}_{1}(\T-2\delta,\T)}\nonumber\\
&=\frac{1}{r^{d+1}_{1}(0,\T)}=\frac{1}{r^{d+1}_{P_1}(\T)}.
\end{align}
It is noted that we have $\frac{{d{r_1}\left( {v ,\T} \right)}}{d\T}=-\frac{{d{r_1}\left( {v ,\T} \right)}}{{dv}}$.\\
For the null boundary in region $i\,\left( {1 < i \le n} \right)$, one can write
\begin{equation}\label{null}
  \int_v^{{\T_{i - 2}}} {dv}  =  - 2\int_{{r_i}\left( {v,\T} \right)}^{{r_{{P_{i - 1}}}}\left( \T \right)} {\frac{{dr}}{{{f_i}\left( r \right)}}}  \Rightarrow \frac{{v - {\T_{i - 2}}}}{2} = \int_{{r_i}\left( {v,\T} \right)}^{{r_{{P_{i - 1}}}}\left( \T \right)} {\frac{{dr}}{{{f_i}\left( r \right)}}} .
\end{equation}
Taking the time derivative of Eq. \eqref{null} results in
\begin{equation}\label{null-dt}
  \frac{{d{r_i}\left( {v ,\T} \right)}}{{d\T}} = \frac{{{f_i}\left( {{r_i}} \right)}}{{{f_i}\left( {{r_{{P_{i - 1}}}}} \right)}}\frac{{d{r_{{P_{i - 1}}}}\left( \T \right)}}{{d\T}}
\end{equation}
and setting $v  = {\T_{i - 1}}$ in the above expression leads to
\begin{equation}\label{dtp}
  \frac{{d{r_{{P_i}}}}}{{d\T}} = \frac{{{f_i}\left( {{r_{{P_i}}}} \right)}}{{{f_i}\left( {{r_{{P_{i - 1}}}}} \right)}}\frac{{d{r_{{P_{i - 1}}}}}}{{d\T}}.
\end{equation}
Moreover, it can be readily checked that
\begin{equation}\label{dtr}
  \frac{{d{r_i}\left( {v ,\T} \right)}}{{d\T}} = \frac{{ - 2}}{{{f_i}\left( {{r_{{P_i}}}} \right)}}\frac{{d{r_{{P_i}}}\left( \T \right)}}{{d\T}}\frac{{d{r_i}\left( {v ,\T} \right)}}{{dv }}.
\end{equation}

\bibliographystyle{utcaps}
\bibliography{all}
\end{document}